\documentclass[10pt, conference]{IEEEtran}
\ifCLASSINFOpdf
\else
\fi

\usepackage{times,amsmath,epsfig} 

\newtheorem{example}{Example}[section]
\newtheorem{mydef}{\textbf{Definition}}

\usepackage{balance}
\usepackage{subfigure}%
\usepackage{graphicx}
\usepackage{epstopdf}
\usepackage{textpos}

\begin{document}

%
% paper title
% can use linebreaks \\ within to get better formatting as desired
\title{Risks of Friendships on Social Networks}

\author{
% author names are typeset in 11pt, which is the default size in the author block
{Cuneyt Gurcan Akcora, Barbara Carminati, Elena Ferrari }%
% add some space between author names and affils
\vspace{1.6mm}\\
\fontsize{10}{10}\selectfont\itshape
DISTA, Universit\`{a} degli Studi dell'Insubria\\
Via Mazzini 5, Varese, Italy\\
\fontsize{9}{9}\selectfont\ttfamily\upshape
\{cuneyt.akcora, elena.ferrari, barbara.carminati\}@uninsubria.it\\
}

% make the title area
\maketitle

\begin{textblock*}{500mm}(.15\textwidth,-6cm)
To Appear in the 2012 IEEE International Conference on Data Mining (ICDM).
\end{textblock*}
\begin{abstract}

In this paper, we explore the risks of friends in social networks  caused by their friendship patterns, by  using real life social network data and starting from a previously defined  risk model. Particularly, we observe that risks of friendships can be mined by analyzing users' attitude towards friends of friends. This allows us to give new insights into  friendship and risk dynamics on social networks.

%Friends on social networks have conflicting roles, as enriching our experience by introducing new people to us may lead to personal information disclosure to unwanted audiences. In this paper we explore the risks of friends in social networks  caused by their friendship patterns. Using real life social network data, we create a risk model to evaluate friends based on their friendships. Particularly, we observe characteristics of friends of friends and how these appeal to user when making privacy choices. We find that risks of friendships can be found by analyzing users' attitude towards friends of friends. We conclude with a discussion of how friends can be labeled in terms of risk, and how these risk labels can be exploited to understand friendship dynamics on social networks.

\end{abstract}

\section{Introduction}
\label{sec:Introduction}

Users register on social networks to keep in touch with friends, as well as to meet with new people. Research works have shown that a big majority of people that we meet online and add as friends are not random social network users; these people are introduced into our social graph by friends \cite{akcora2011network}. %,liben2007link
 Although friends can enrich the social graph of users, they can also be a source of privacy risk, because a new relationship  always implies the release of some personal information to the new friend as well as to friends of the new friend, which are {\em strangers} for the user. This problem is aggravated  by the fact that users can reference resources of other users in their social graph; and make it very difficult to control the resources published by a user. This uncontrolled information flow highlights the fact that creating a new relationship might expose users to some privacy risks.  
 
We cannot assume that friends will make the right choices about friendships, because friends may have a different view on people  they want to be friends with. Considering this, privacy of a social network user should be protected by building a model that observes friendship choices of friends, and assigns a risk label to friends accordingly. Such a model requires knowing a user's perception on the risks of friends of friends.  We made a first effort in this direction  in \cite{akcoraicde} by proposing a risk model to learn risk labels of  strangers  by considering  several dimensions. To validate the model, we developed  a browser extension  showing for each stranger (i) his/her profile features, (ii) his/her privacy settings,  and (iii) mutual friends. Based on this information, the user is asked to give a risk label $l\in\{1,2,3\}$ to the stranger. These risk labels correspond to \emph{not risky, risky} and \emph{very risky} classification of a stranger. Through the extension, 47 users (32 male, 15 female users) have labeled 4013 strangers. However, {\em we did not consider risk of friends}. 

This new work starts with considering two factors in assigned risk labels.  First,  strangers can be risky only because of their profile features. Second, a friend himself can increase or decrease the risk of a stranger. Increases and decreases will be termed as {\em negative and positive friend impacts}, respectively. In any case, if a risky stranger is introduced into the user's social graph  it is because of his/her friendship with a friend. However, determining the friend impacts can help us to  determine which privacy actions should be taken to avoid data disclosure.  We aim at learning how  risk labels are assigned to strangers depending only on their profile features, and how much a friend can impact (i.e., increase or decrease) these labels. If strangers  are risky just because of their profile features, privacy settings can be restricted to avoid only these strangers. On the other hand, if a friend increases the risk labels of strangers, all of his/her strangers should be avoided.

We begin our discussion with reviewing the related work in Section \ref{sec:RelatedWork}. In Section \ref{sec:ModelingTheRiskOfFriendships} we explain the building blocks of our model and Section \ref{sec:TransformingData} shows  how we use our dataset efficiently. In Section \ref{sec:BaselineEstimation} we discuss the role profile features in risk labels, and in Section \ref{sec:FriendImpact} we show how impacts of friends are modeled. Section \ref{sec:FriendRisklabels} explains finding risk labels of friends from friend impacts, and in Section \ref{sec:ExperimentalResults} we give the experimental results.

\section{Related Work}
\label{sec:RelatedWork}

Friends' role in user interactions has been studied in sociology \cite{rubin1985just}, but observing it on a wide scale has not been possible until online social networks attracted millions of users and provided researchers with social network data. For online social networks, Ellison et al. \cite{Ellison2007} defined friends as social capital in terms of an individual's ability to stay connected with members of a previously inhabited community. Differing from this work, we study how friends can help users to interact with new people on social networks. Although these interactions can increase users' contributions to the network \cite{stutzman2010friends} and help the social network evolve by creation of new friendships \cite{valafar2009beyond}, they can also impact the privacy of users by disclosing profile data. Squicciarini et al. \cite{squicciarini2010prima} have addressed concerns of data disclosure by defining access rules that are tailored by 1) the users' privacy preferences, 2) sensitivity of profile data and 3) the objective risk of disclosing this data to other users. Similarly, Terzi et al. \cite{liu2009framework} has considered the sensitivity of data to compute a privacy score for users. Although these works regulate profile data disclosure during user interactions, they do not study the role of friends who connect users on the social network graph and facilitate interactions.  Indeed, research works (see \cite{so74094} for a review) have been limited to finding the best privacy settings by observing the interaction intensity of user-friend pairs \cite{banks2009all} or by asking the user to choose privacy settings \cite{fang2010privacy}. Without explicit user involvement, Leskovec et al. \cite{Leskovec2010pos} have shown that the attitude of a user toward another can be estimated from evidence provided by their relationships with other members of the social network. Similar works try to find friendship levels of two social network users (see \cite{waqar2010} for a survey). Although these work can explain relations between social network users, they cannot show how existence of mutual friends can change these relations. 

Privacy risks that are associated by friends' actions in information disclosure has been studied in  \cite{thomas2010unfriendly}, but the authors work with direct actions (e.g., re-sharing user's photos) of friends, rather than their friendship patterns. 
Recent privacy research focused on creating global models of risk or privacy rather than finding the best privacy settings, so that ideal privacy settings can be mined automatically and presented to the user more easily. In \cite{akcoraicde}, Akcora et al. prepared a risk model for social network users in order to regulate personal data disclosure. Similarly, Terzi et al. \cite{liu2009framework} has modeled privacy by considering how sensitive personal data is disclosed in interactions. Although users assign global privacy or risk scores to other social network users, friend roles in information disclosure are ignored in these work. 

An advantage of global models is that once they are learned, privacy settings can be transfered and applied to other users. In such a shared privacy work, Bonneau et al. \cite{bonneau2009privacy} use \textit{suites} of privacy settings which are specified by friends or trusted experts. However, the authors do not use a global risk/privacy model, and users should know which suites to use without knowing the risk of social network users surrounding him/her.

\section{Overall Approach}
\label{sec:ModelingTheRiskOfFriendships}

We will start this section by explaining the terminology that will be used in the paper. In what follows, on a social graph $\mathcal{G}_u$, 1 hop distance nodes from $u$ are called friends of $u$, and 2 hop distance nodes are called strangers of $u$, i.e., strangers of user $u$ are friends of friends of $u$. We will denote all strangers of user $u$ with $S_u$, and risk label of each stranger $s \in S_u$ that was labeled by $u$ will be denoted as $l_{us}\in\{1,2,3\}$. 

A social network $\mathcal{G}=(N, E, Profiles)$ is a collection of $N$ nodes and $E \subseteq N \times N$ undirected edges. $Profiles$ is a set of profiles, one for each node $n \in \{ 1,...,\left|N\right|\}$. A social graph $\mathcal{G}_u=(V, R, F)$ is constructed from the social network $\mathcal{G}$ for each user $u \in N$, such that, the node set $V = \{\forall n \in N |distance(n,u)\leq 2\}$. Nodes in $\mathcal{G}_u$ consist of friends and strangers of $u$. Similarly, edge set R consists of all edges in $\mathcal{G}$ among nodes in $V$. Each node $v \in V$ in a social graph will be associated with a feature vector $f_v \in F$. Cells of $f_v$ correspond to profile feature values from the associated user profile in $Profiles$.

The goal of our model is to assign risk labels to friends according to the risk labels of their friends (i.e., strangers). As we stated before, risk labels of strangers depend on stranger features as well as mutual friends \cite{akcoraicde}. We do not assume that all friends can change users' risk perception in the same way. Some friends can make strangers look less risky and facilitate interactions with them (i.e., friends decrease the risk of strangers). On the other hand, some friends can make strangers more risky (i.e., friends increase the risk of strangers). For example, if users do not want to interact with some friends, they might avoid friends of these friends as well. We will use positive and negative impacts to refer to decreases and increases in stranger risk labels, respectively. To understand whether friends have negative or positive impacts, our model must be able to know what risk label the stranger would receive from the user if there were no mutual friends. This corresponds to the case where the user given label depends only on stranger features. We will term this projected label as the baseline label, and show it with $b_{us}$. For instance, assume that if there are no mutual friends, a user $u$ considers all male users as very risky, and avoids interacting with them. In this case, the baseline label for a male stranger $s$ is very risky, i.e.,  $b_{us}=very~risky$. However, if the same male stranger $s$ has a mutual friend with user $u$, we assume that the user given label $l_{us}$ might not be equal to the baseline label $b_{us}$ (i.e.,$l_{us} \neq b_{us}$), because the mutual friend might  increase or decrease the risk perception of the user. This difference between the baseline and user given labels will be used to find out friend impacts.

Finding baseline labels and friend impacts requires different approaches. In baseline estimates, we use logistic regression on stranger features, and for the friend impacts we use multiple linear regression \cite{myers1990classical}. Both of these regression techniques require many user given labels to compute baseline labels and friend impacts with high confidence.  However, users are reluctant to label many strangers, therefore we have to exploit few labels to achieve better results. To this end, we transform our risk dataset, and use the resulting dataset in regression analyses. In the next sections, this transformation and regression steps will be described in detail. Overall, we divide our work into four phases as follows:

\begin{enumerate}
	\item \textbf{Transformation:} Exploit the risk label dataset in such a way that regression analyses for baseline labels and friend impacts can find results with high confidence. With this step, we increase the number of labels that can be used to estimate baseline labels and friend impacts.
	\item \textbf{Baseline Estimation:} Find baseline labels of strangers by logistic regression analysis of their features.
	\item \textbf{Learning Friend Impacts:} Create a multiple linear regression model to find friends that can change users' opinion about strangers and result in a different stranger label than the one found by baseline estimation.
	\item \textbf{Assigning Risk Labels to Friends:} Analyze the sign of friend impacts, and assign higher risk labels to friends who have negative impacts.
	
\end{enumerate}

\section{Transforming Data}
\label{sec:TransformingData}

By transforming the data, we aim at using the available data efficiently to find friend impacts with higher confidence. To this end, we first transform profile features of friends and strangers to use \textit{k-means} and hierarchical clustering algorithms \cite{gan2007data} on the resulting profile data. This section will discuss the transformation, and briefly explain the clustering algorithms.

Our model has to work with few stranger labels, because users are reluctant to label many strangers. This limitation is also shared in Recommender Systems (RS) \cite{melville2010recommender} where the goal is to predict ratings for items with minimum number of past ratings. In neighborhood based RS \cite{koren2010factor}, ratings of other similar users are exploited to predict ratings for a specific user. Traditionally, the definition of similarity depend on the characteristics of data (e.g., ordinal or categorical data), and it has to be chosen carefully. We use profile data of friends and strangers in defining similar friends and strangers, respectively. Friend impacts of a user $u$ is learned from impacts of similar friends from all other users. To this end, we transform  profile data of friends and strangers in such a way that friends and strangers of different users are clustered into global friend and stranger clusters. Next sections will describe the aims and methods of friend and stranger clustering in detail.  

\subsection{Clustering Friends}
\label{sec:ClusteringFriends}

Clustering friends aim at learning friend impacts for a cluster of friends. This is because we might not have enough stranger labels to learn impacts of individual friends with high confidence. To overcome this data disadvantage, impact of a friend $f$ can be used to find the impacts of other friends who belong to the same cluster. For example, a user from Milano can have a friend from Milano, whereas a user from Berlin can have a friend from Berlin. Although these two friends have different hometown values (Milano and Berlin), we can assume that both friends can be clustered together because their hometown feature values are similar to user values. This hometown example demonstrates a clustering based on a single friend profile feature and it results in only two clusters: friends who are from Milano/Berlin and friends who are from somewhere else. However, in real life social networks, friends have many values for a feature, some of which can be more similar to the user's value than others. For example, Italian friends of a  user from Milano can be from Italian cities other than Milano, and these friends should not be considered as dissimilar as friends from Berlin. By considering these, we transform categorical friend values to numerical values in such a way that similarities between friend and user values become more accurate.

Our transformation uses the homophily \cite{mcpherson2001birds} assumption which states that people create friendships with other people who are similar to them along profile features such as gender, education etc. In other words, we assume that all friends of a user $u$ can be used to judge the similarity of a social network user to $u$. For example, considering the case where the user $u$ is from Milano, a social network user from Rome is similar to the user if the user has many friends from Rome. Moreover, we assume that different users will have similar clusters of friends, e.g., friends from user's hometown, alma mater etc. and friend impact values will be correlated with their corresponding clusters, e.g., friends from hometowns will have similar impact values. More precisely, the transformation of friends' data maps a categorical feature value of a friend, such as hometown:Milano, to a numerical value which is equal to the frequency of the feature value among profiles of all friends of a user. For example, if a friend $f$ has profile feature value hometown:Milano, and there are 15 out of 100 friends with similar hometown:Milano values, hometown feature of $f$ will be represented with $15/100=0.15$. After applying this numerical transformation to all friends of all users, we compute a Social Frequency Matrix  for Friends (SFMF) where each row represents numerical transformation of feature vector of a user's friend.

\begin{mydef}[Social Freq. Matrix for friends]

The Social Frequency Matrix associated with a social network $\mathcal{G}$ is defined as $|N| \times |F| \times n$, where N is the set of users in $\mathcal{G}$, $F\subset N$ is the set of user in $\mathcal{G}$ that are friends of at least one  user $u\in N$, and n is the number of features of user profiles. Each element value of the matrix is given by: 

$$SFMF[u,f,v]=\frac{Sup({\vec{f}_v})}{ \left | F_u \right |} $$ where $F_u\subset F$ is the set of  friends of $u$,  $Sup(\vec{f}_v)=\left|\{g\in F_u|\vec{g}_v=\vec{f}_v\}\right|$ and $f\in F_u$, whereas $\vec{g}_v$ and  $\vec{f}_v$ show the value of profile feature $v$ for  users $g$ and $f$, respectively.

\end{mydef}

Having transformed friend data into numerical form, we can now use a clustering algorithm to create clusters of friends. After applying a clustering algorithm to the Social Frequency Matrix for friends, output friend clusters will be denoted by $FC$. 

\subsection{Clustering Strangers}
\label{sec:ClusteringStrangers}

By clustering friends, we can learn impacts of friends from different clusters, but this raises another question: do friends have impact on all strangers of users? Our assumption is that correlation between stranger and friend profile features can reduce or increase friend impact. For example, if a student user $u$ labels friends of a classmate friend $f$, we might expect friends of $f$ who are professors to have higher risk labels than student friends of $f$, because $u$ might not want his/her professors to see his/her activities and photos. Here the work feature of strangers changes friend impact of $f$ by increasing the risk label of professor friends of $f$. To see how friend and stranger features change friend impacts, we transform strangers' profile data to numerical data and cluster the resulting matrix just like we clustered friends.  This clustered stranger representation helps us detect clusters of strangers for whom certain clusters of friends can change risk perception of users the most. Formally, we prepare a social frequency matrix as follows:

\begin{mydef}[Social Freq. Matrix for strangers]

The Social Frequency Matrix for Strangers associated with a social network $\mathcal{G}$ is defined as $|N| \times |S| \times n$, where N is the set of users in $\mathcal{G}$, $S\subset N$ is the set of user in $\mathcal{G}$ that are strangers of at least one  user $u\in N$, and n is the number of features of user profiles. Each element value of the matrix is given by: 

$$SFMS[u,s,v]=\frac{Sup({\vec{s}_v})}{ \left | F_u \right |} $$ where $Sup(\vec{s}_v)=\left|\{g\in F_u|\vec{g}_v=\vec{s}_v\}\right|$ and $S\in N$, whereas $\vec{s}_v$ shows the value of feature v for  stranger $s$.

\end{mydef}

Note that we still use friend profiles in the denominator to transform stranger data. This is because we cannot see all strangers of a friend due to API limitations of popular social networks. To overcome this problem, we use friend profiles because we expect them to be similar to profiles of their own friends (strangers). We again use the Social Frequency Matrix for strangers to create clusters of strangers. We will denote stranger these stranger clusters by $SC$.

\subsection{Clustering Algorithms}
\label{sec:ClusteringAlgorithms}

In our experiments, we used the \textit{k-means} and hierarchical algorithms \cite{gan2007data} to produce clusters of friends and strangers. This section will briefly explain these algorithms. In what follows, we will use data points and strangers/friends interchangeably to mean elements in a cluster.

The \textit{k-means} clustering algorithm takes the number of final clusters as input and clusters the data by successively choosing cluster seeds and refining the distance within cluster data points. The required input for the number of final clusters is usually unknown beforehand and this makes \textit{k-means} unfeasible in some scenarios. However, in our model it gives us the flexibility to experiment with different sizes of clusters. \textit{k-means} is also a fast clustering algorithm which suits our model for the cases where all friends of all users can reach a few thousands. In our experiments, we used different $k$ values to find optimal performance. In hierarchical clustering\footnote{We used the agglomerative form where a new stranger is added to clusters by considering the complete distance. Height of the tree was 3.}, a tree structure is formed by joining clusters and the tree is cut horizontally at some level to produce a number of clusters. 

In friend and stranger clustering, choosing the number of final clusters or the horizontal level requires some trade-offs. The advantage of using many clusters is that data points in each cluster are more similar to each other (i.e., friends or strangers in a cluster are more similar in profile feature values). On the other hand, too many clusters decreases the average number of data points in a cluster, and our model may not be trained on these clusters with high confidence, i.e., there may not be enough data points in a cluster to prove anything. Using too few clusters also has a disadvantage. Final clusters may contain too many data points that are not very similar to each other. This decreases the quality of inferences because what we infer from some data points might not be valid for others in the same cluster. Despite this, if data points are naturally homogeneous, the similarity among data points in a big cluster can be high. As a result, a big cluster may offer more data to prove our inferences with more confidence.

After transforming our data and creating friend and stranger clusters, we will now explain baseline label estimation for stranger clusters.

\section{Baseline Estimation}
\label{sec:BaselineEstimation}

Baseline estimation analyzes how feature values on stranger profiles bring users to assign specific risk labels to strangers. The baseline estimation process results in baseline labels for each stranger $s \in S$. These labels are found by using statistical regression methods on already given user labels and stranger profile features. In this section we will discuss this process.

\begin{figure}[t]
	\centering
		\includegraphics[width=1.0in]{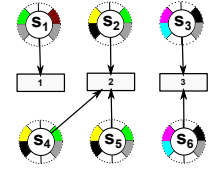}
	\caption{\small Features and risk labels \label{fig:features}}
\end{figure}

Baseline estimation corresponds to the case where a user would assign a risk label to a stranger without knowing which one of his/her friends are also friends with the stranger. Figure \ref{fig:features} shows an example of baseline estimation. In the figure, each stranger $s \in S_u$ is a node surrounded by a ring representing his/her feature vector $f_s$. Each cell in the feature vector corresponds to a feature value of the stranger (e.g., hometown:Milano). Different colors for the same cell position represent different values for the same feature on different stranger profiles. In the example shown in Figure \ref{fig:features} strangers $S_2$, $S_4$ and $S_5$ are labeled with 2 (i.e., the risky label). These three strangers share the same feature vector as shown with the same colored cells. Based on these observations, if any stranger  has the same feature vector with $S_2$, $S_4$ and $S_5$, the stranger will be given label 2. The evidence to support this statement comes from the three strangers ($S_2$, $S_4$ and $S_5$), and the number of such strangers determine the confidence of the system in assigning baseline labels.

Although in Figure \ref{fig:features} stranger features are shown to be the only parameter in defining stranger labels, in our dataset labels of strangers have been collected from users by explicitly showing at least one mutual friend in addition to the stranger feature values. Because of this, stranger labels that are learned from users can be different from baseline labels; they can be higher (more risky) or lower (less risky) depending on the friend impact. Considering this, in baseline estimation we use the labels of strangers who have the least number of mutual friends with users. These are the subset of labels which were given to strangers who have only one friend in common with users, i.e., for user $u$ and stranger $s$, $\left|F_u \cap F_s\right|=1$. In what follows, we will use \textit{first group dataset} to refer to these strangers.

In our approach, we use logistic regression to learn the baseline labels from available data. This allows us to work with categorical response variables (i.e., one of the tree risk labels). Stranger features are used as explanatory (independent) variables and risk labels as the response (dependent) variable which is determined by values of explanatory variables (i.e., feature values). Although the response variable has categorical values, it can be considered  ordinal because risk labels can be ordered as not risky (label 1), risky (label 2) and very risky (label 3).

Ordinary Logistic Regression is used to model cases with binary response values, such as 1 (a specific event happens) or 0 (that specific event does not happen), whereas multinomial logistic regression is used when there are more than two response values. As multinomial logistic regression a basic variant of logistic regression, we will first start with the definition of logistic regression. For this purpose, assume that our three risk labels are reduced to two (risky, not risky).

Suppose that $\pi$ represents the probability of a particular outcome, such as a stranger being labeled with risky, given his/her profile features as a set of explanatory variables $x_1,...,x_n$:
$$P(l=risky)=\pi=\frac{e^{(\alpha+\sum{ \beta_k  X_k })}}{1+e^{(\alpha+\sum{\beta_k X_k})}}$$ 

where $0 \leq \pi \leq 1$, $X_k$ is a feature value, $\alpha$ is an intercept and $\beta$s are feature coefficients, i.e., weights for feature values. The logit transformation $log[\frac{\pi}{1-\pi}]$ is used to linearize the regression model:
$$log[\frac{\pi}{1-\pi}] = \alpha+\sum{ \beta_k  X_k }$$

By transforming the probability ($ \pi $) of the response variable to an odd-ratio ($log[\frac{\pi}{1-\pi}]$), we can now use a linear model. Given the already known stranger features and labels, we use Maximum Likelihood Estimation \cite{myung2003tutorial} to learn the intercept value and the coefficients of all features.

Although standard binary logistic regression and multinomial logistic regression use the same definition, they differ in one aspect: multinomial regression chooses a reference category and works with not one but $N-1$ log odds where $N$ is the number of response categories. In our model, $N=3$, because the response has three labels (1, 2, 3). In both binary and multinomial logistic regression, intercept and coefficient values are found by using numerical methods to solve the linearized equation(s). With the found values, we can write the odd ratio as an equation. For example, in equation  $log[\frac{\pi}{1-\pi}] = 0.7+ 1.2 \times X_1+0.3 \times X_2$, the intercept value is (0.7) and feature coefficients ($\beta_1=1.2$ and $\beta_2=0.3$). We can then plug in a new set of values (e.g., $X_1=0.5$) for features, and get the probabilities of response value being one of three labels. For example, for a specific stranger, the model can tell us that risk label probabilities of the stranger is distributed as \%0.9 very risky, \%0.09 risky and \%0.01 not risky. As we can compute baseline label in real values, a stranger $s \in S$ is assigned a baseline label by weight averaging the probabilities of risk labels. 

\section{Friend Impact}
\label{sec:FriendImpact}

So far, we have discussed clustering and baseline label estimation. In this section we will first discuss how these two aspects of our model are combined to compute friend impacts.  After finding friend impacts, we will discuss how risk labels can be assigned to friends by considering the sign of impact values. In computing friend impacts, we use multiple linear regression \cite{myers1990classical}, which learns friend impacts by comparing baseline and user given labels to strangers. To this end, we define an estimated label parameter to use in linear regression as follows:

\begin{mydef}[Estimated label]
\label{frienimpact}
For a stranger $s$ and a user $u$, an estimated label is defined as:
$$ \hat{l}_{us}= b_{us}+ \sum \limits_{FC_i \in FC}{ FI(FC_i,SC_j)}  \times Past(u,s)   $$
\end{mydef}

where $\hat{l}_{us}$ and $b_{us}$ are estimated and baseline labels for a stranger $s$, and $s$ belongs to the stranger cluster $SC_j \in SC$. Friend clusters  $FC$ are found by applying a algorithm to the mutual friends of user $u$ and stranger $s$. $Past(u,s)$ denotes an intermediary value based on stranger labels given by user $u$,  whereas $FI(FC_i,SC_j)$ represents impact of a friend $f$ from a friend cluster $FC_i$  on the label of stranger $s$ from a stranger cluster $SC_j$.

In the rest of this section, we will define the $Past(.,.)$ and $FI(.,.)$ parameters, and explain how they are used to compute friend impacts. 

\subsection{The Past Labeling Parameter}
\label{sec:Defining}

We start by discussing the past parameter $ Past(.,.)$ which returns a value from past labellings of strangers by user $u$. 

The past parameter is traditionally used in recommender systems to adjust baseline estimate \cite{melville2010recommender}. The need for this parameter arises from the fact that baseline estimation is computed from labels of all strangers who have only one mutual friend with user $u$ (i.e., first group dataset), and it tends to be a rough average. To overcome this, a subset of strangers, who are very \emph{similar} to $s$ and who have been labeled in the past by $u$, are observed and the baseline label is increased or decreased to make it more similar to the user given labels of these strangers.

In defining the past parameter, we consider two factors: how many similar strangers should be considered in this adjustment and what is an accurate metric for finding similarity of two strangers? For the first question, we use the computed stranger clusters. For a stranger $s$, similar strangers from the first group dataset are those (i) that are labeled by the same user $u$,  and (ii) that belong to the same  stranger cluster with $s$. Although we use stranger clusters to choose similar strangers, the similarity of strangers in a cluster can be low or high depending on the clustering process. With too few clusters and too many clusters, similarity of strangers in a cluster can be low and high respectively. We adjust the baseline labels by considering labels given to most similar users. To this end, we use the profile similarity measure by Akcora et al. \cite{akcora2011network}. This measure assigns a similarity value of 1 to strangers with identical profiles, and for non-identical profiles the similarity value is higher for strangers whose profile feature values are more common in profile features of $u$'s friends. Formally, we define the past labeling as follows:

\begin{mydef} [Past Labeling Parameter]$\hfill$

For a given user $u$ and stranger $s$, the past labeling parameter is defined as:  

$$Past(u,s)=\frac{1}{\left|SC_i\right|}\sum\limits_{x \in C_i}{PS(s,x) \times (l_{ux} - b_{ux})} $$ where $PS()$ denotes the profile similarity between two strangers, $l_{ux}$ is the user given label of stranger x, and $b_{ux}$ is the baseline label of x. Strangers $s$ and $x$ belong to the same stranger cluster $C_i$.
\end{mydef}

\subsection{The Friend Impact Parameter}
\label{sec:TheFriendImpacts}

The second parameter from Definition \ref{frienimpact}, $FI(f,s)$, is used to show impacts of mutual friends on the risk label given to $s$ by $u$. In modeling friend impacts, we wanted to see how friends from different clusters changed the baseline label. By using this approach, we explain impacts of friend clusters in terms of friend features that shape friend clusters. If there is at least one mutual friend from a friend cluster $FC_i$,  we say that friend cluster $FC_i$ may have impacted the label given to the stranger $s$. For the cases where a stranger $s$ has two or more mutual friends from a friend cluster $FC_i$, we experimented with both options for $FI(f,s)$. Next, we will explain these options.

\begin{figure}
\centering
\subfigure[Multiple impacts for a friend cluster.]{
\includegraphics[scale=0.5]{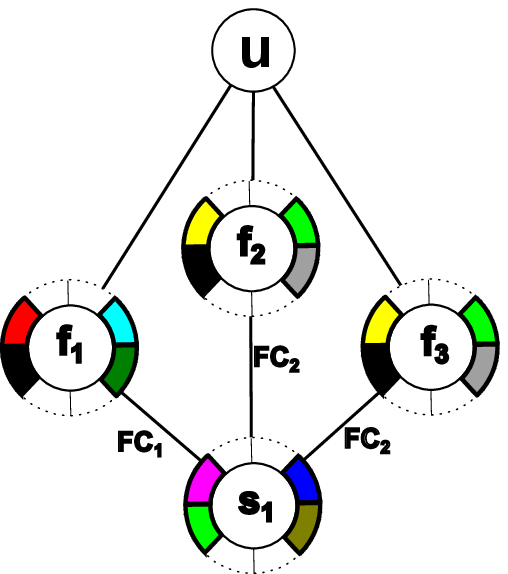}
\label{fig:single}
} \hspace{2mm}
\subfigure[Single impact for a friend cluster.]{
\includegraphics[scale=0.5]{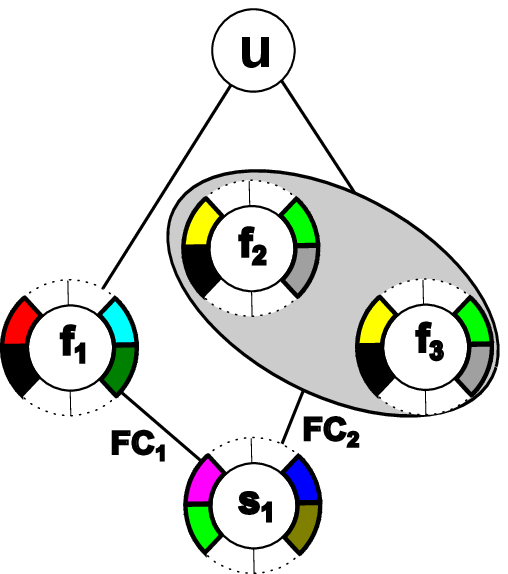}
\label{fig:combined}
}
\caption{Friend impact definitions by considering the number of friends from the same cluster. In the single impact definition, two friends do not increase the friend impact.}
\label{fig:imp}
\end{figure}

\subsubsection{Multiple Impact for the Friend Cluster}
\label{sec:MultipleFriendImpact}

In our first approach, we assume that a bigger number of mutual friends from friend cluster $FC_i \in FC$ will impact user labeling. Assume that from a friend cluster $FC_i \in FC$, we are given a set of mutual friends $MF_i=\{\forall f |f \in FC_i, f \in \{F_u \cap F_s\}\}$ of user $u$ and stranger $s$. We define the impact of friend cluster $FC_i$ on the label of stranger $s\in SC_j$ as follows: 

$$FI_2(FC_i,SC_j) =  \left | MF_i \right | \times I_{FC_i,SC_j}$$ where $I_{FC_i,SC_j}$ is the impact of a cluster $FC_i|f \in \{FC_i \cap MF_i\}$ on the label of stranger $s\in SC_j$. Note that this impact ($I_{FC_i,SC_j}$) is the unknown value that our system will learn.

\subsubsection{Single Impact for the Friend Cluster}
\label{sec:SingleFriendImpact}

In the second approach, we assume that a bigger number of friends from the same cluster does not make a difference in user labeling; at least one friend from the cluster is required, but more friends do not bring additional impact. This approach is shown in Figure \ref{fig:combined}, where friends are shown with their cluster ids, and two friends from friend cluster $FC_2$ bring a single impact. Assume that from a friend cluster $FC_i \in FC$, we are given a set of mutual friends of user $u$ and stranger $s$. We give the impact of friend cluster $FC_i$ on the label of stranger $s$ as follows: 

$$FI_1(FC_i,SC_j) =  I_{FC_i,SC_j}$$ where $I_{FC_i,SC_j}$ is the impact of a friend cluster $FC_i$ on label of stranger $s \in SC_j$.

These different friend impact approaches change the model by including different numbers of friend impacts. The unknown impact variable $I_{FC_*,SC_*}$ is learned by the least squares method \cite{jiang1998least}. The least squares method provides an approximate solution when there are more equations than unknown variables. In our model, each stranger's label provides an equation to compute impacts of $k_1$ friend clusters on $k_2$ stranger clusters ($k_1$ and $k_2$ are the final numbers of friend and stranger clusters in the k-means algorithm). In Example \ref{ex:friendimpact}, we will explain these points and give equations of one stranger for single and multiple impact definitions.

\begin{example}
\label{ex:friendimpact}
Given a stranger $s_1\in SC_1$ who is labeled by $u$, assume that the user given label $l_{us_1}=2.3$, while the baseline label is $b_{us_1}=2.7$. Again assume that $Past(u,s)=-0.2$. Equations for the stranger $s$ with single and multiple friend impact definitions are respectively given as follows:

$$2.3=2.7+( I_{FC_2,SC_1}+I_{FC_1,SC_1})\times -0.2$$
$$2.3=2.7+( 2\times I_{FC_2,SC_1}+I_{FC_1,SC_1})\times -0.2$$
\end{example}

After choosing one of these definitions of friend impact, we input one equation for each stranger $s$ to the least squares method to compute impact values of friend clusters on stranger clusters. In the experimental results, we will discuss the definition that yielded the best results.

\section{Friend Risk Labels}
\label{sec:FriendRisklabels}
Learning impact values allows us to see the percentage of positive and negative impacts for each friend cluster. Negative impact values for a friend cluster shows that the friend cluster increases the risk label of strangers. Depending on a user's choice, friend clusters which have negative impacts less than $x\%$ of the time can be considered not risky. Similarly, a threshold $y\%$ can be chosen to determine very risky friend clusters. In our experiments, we heuristically chose $x=20$ and $y=50$. With these threshold values for risk labels, we formally define the risk label of a friend $f$ as follows:
\begin{mydef}[Friend Risk Label]
Assume that the percentage of positive and negative impact values for a cluster $FC_i \in FC$ are denoted with $Im_{i}^+$ and $Im_{i}^-$ respectively, where $Im_{i}^+ + Im_{i}^- =1$. We assign a risk label to a friend $f$ who is a member of the friend cluster $FC_i$ (i.e., $f \in FC_i$) according to the negative impact percentage of the friend cluster $FC_i$ as follows:
\[
l(u,f) = \left\{
\begin{array}{l l}
\text{not risky} & \quad \text{if} \quad Im_{i}^{-} < 0.2 \\
\text{risky} & \quad \text{if}\quad 0.2 \leq Im_{i}^{-}	< 0.5\\
\text{very risky} & \quad \text{if}\quad Im_{i}^{-}	\geq 0.5 \\\end{array} \right \}
\]
\end{mydef}

Next we will give the experimental results of our model performance.
\section{Experimental Results}
\label{sec:ExperimentalResults}
In this section we will validate our model assumptions, and then continue to give detailed analysis of performance under different parameter/setting scenarios.

\subsection{Validating Model Assumptions}
\label{sec:ValidatingModelAssumptions}

Before finding friend impacts, we validated our model assumption (i.e., mutual friends have an impact on the risk label of a stranger) by using logistic regression on the whole dataset (4013 stranger labels and profiles). For this, we included \textit{the number of mutual friends} as a parameter, and computed the \textit{significance}\footnote{Significance is measured by p-values. The p-value is the probability of having a result at least as extreme as the one that was actually observed in the sample. Traditionally, a $p-value$ of less than 0.05 is considered significant.} of model parameters. In overall regression, photo visibility, wall visibility, education and work parameters were excluded from the model because they were found to be non-significant. For significant parameters, $Pr(>\left|t\right|)$ values are shown in Table \ref{table:allregresults}. 

In the regression, there are two friend related parameters: the number of mutual friends and the friendlist visibility. Differing from the number of mutual friends, friendlist visibility is a categorical variable which takes 0 when the stranger hides his/her friendlist from the user and 1 otherwise. From Table \ref{table:allregresults}\footnote{	Notes: Reference category for the equation is label 2. Standard errors in parentheses. Significance codes:  '***' 0.001 '**' 0.01 '*' 0.05 '.' 0.1 ' ' 1 }, we see that seeing a stranger's friendlist increases the probability of the stranger getting label 1, whereas it is not an important parameter for label 3.  Our main focus in regression analysis was to verify that the number of mutual friends parameter is significant. We found that an increasing number of mutual friends indeed helps a stranger get label 1, and decreases the probability of getting label 3. This result tells us that friends have an impact on user decisions and our assumption about the existence of friend impacts holds true. After validating our model assumption, we continue to the baseline label estimations.

\begin{table}
\caption{\small{ Regression results for all data points. p-value=2.22e-16. Total N=4013. 
		}}
	\label{table:allregresults}
\centering
	
	\begin{small}
  \begin{tabular}{l l l}
 				\textit{ } & Label 1 & Label 3\\ 
			
				\small{ Intercept} &\small{-1.2668850***} &\small{-0.7626810***} \\
				&(0.146)&(0.138)\\
				\hline
				\small{Mutual friends}& \small{0.0379547***} & \small{-0.0467834***}  \\ 
				&(0.008)&(0.012)\\
				\hline
				\small{Gender}& \small{-0.3696749**} & \small{0.3480055**}  \\ 
				&(0.118)&(0.113)\\
				\hline
				\small{Friendlist visibility}& \small{0.6203365***} & \small{-0.0642952} \\  
				&(0.125)&(0.118)\\
				\hline
				\small{Locale}& \small{0.6167273***}  & \small{0.7070663***}  \\ 
				&(0.180)&(0.172)\\
				\hline
				\small{Location}& \small{0.1347104}  & \small{0.2708697*} \\ 
				&(0.128)&(0.125)\\
				
				N&1116&1161\\
	
		\hline		
				\end{tabular}
		\end{small} 
		
\end{table}

\subsection{Training for Baseline}
\label{sec:BaselineCalculation}

Baseline calculation predicts labels for strangers without friend impacts. For this purpose we take strangers who have one mutual friend with users ($|MF|=1$) into a new dataset (first group dataset), and train a logistic regression model. 
Logistic regression on the first group dataset finds how stranger features bring users to label strangers.  Table \ref{table:firstregresults} shows model parameters and their corresponding $p$-values.

\begin{table}
\caption{\small{Regression results for the first group data points. p-value = 5.6701e-11. Total N=1520. 
	}}
	\label{table:firstregresults}
\centering
	
	\begin{small}
  \begin{tabular}{l c c}
				\textit{ } & Label 1 & Label 3\\ 
				
				Intercept &-2.5400*** &-0.8661*** \\
				&(0.6305)&(0.2791)\\
				\hline
				Gender& -1.1026** & 0.6985*  \\ 
				&(0.4108)&(0.3350)\\
				\hline
				Friendlist visibility& 0.4705* &  0.5214 \\  
				&(0.2075)&(0.1706)\\
				\hline
				Wall& 0.4173\LARGE{.}  & -0.1595  \\ 
				&(0.2463)&(0.2262)\\
				\hline
				Photo&1.9425** & 0.1361\\
				&(0.6093)&(0.2339)\\
				\hline
				Locale & 0.1446 & 0.5846*\\
				&(0.2881)&(0.2277)\\
				
				N&278&588\\
		\hline
		
				\end{tabular}
		\end{small} 
\end{table}

In Table \ref{table:firstregresults}, we see that when users label the first group strangers, photo and wall visibility are significant parameters. If these items are visible on stranger profiles, the probability of strangers getting label 1 increases. In the whole dataset (see Table \ref{table:allregresults}), these two parameters were found to be insignificant. Another interesting result is that locale\footnote{Locale is the web interface language of the user on the social networking site (e.g., IT for Italian and RU for Russian).} is significant for label 3 whereas it is non/significant for label 1. A high locale value means that the stranger is similar to existing friends of users, but this high similarity is shown to increase the probability of strangers being labeled as very risky, i.e., receiving label 3.

\begin{figure}[t]
	\centering
		\epsfig{file=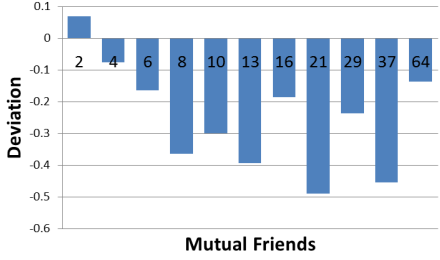, width=0.7\linewidth}
	\caption{\small Deviation of user given labels from baseline labels. Values in the x-axis are the number of mutual friends between a stranger and user. \label{fig:labelvariation}}
\end{figure}

After computing a baseline label for all strangers, we use the difference between user given and baseline labels (${l}_{us}-b_{us}$) to model the friend impact. These differences (deviations from the baseline label) are shown in Figure \ref{fig:labelvariation}. In the figure, we see that user given labels are lower than the computed baseline label, which shows that in overall friends have positive impacts (i.e., thanks to mutual friends, users assign lower risk labels to strangers.). Overall, we found that there  is not a linear relation between the number of mutual friends and the deviation values. This non-linearity changes how we define  the impacts of friend clusters. In Section \ref{sec:FriendImpact} we gave two definitions for friend impacts (see Figure \ref{fig:imp}) to account for deviations from the baseline label. 

In multiple friend impacts we assumed that more mutual friends from a friend cluster bring additional impacts. On the other hand, in single friend impact one friend was enough to have the impact of a friend cluster. This finding implies that more friends of the same cluster do not provide any benefits to strangers on Facebook and mutual friends from different clusters are more suitable to change the user's risk perception about a stranger. We believe that this can be generalized to other undirected social networks.

In the rest of the experiments, we will give the results computed by using the single friend impact definition. We will now explain the model performance under different clustering settings.

\subsection{Clustering}
\label{sec:Clustering}

For clustering 12659 friends, and 4013 strangers we experimented with k-means and hierarchical clustering algorithms. In our experiments with different numbers of final clusters, the k-means algorithm yielded the best results for friend clustering, whereas hierarchical clustering was better for stranger clustering. Due to space limitations, we will omit hierarchical clustering results for friends and k-means results for strangers.

\begin{figure}
	\centering
		\epsfig{file=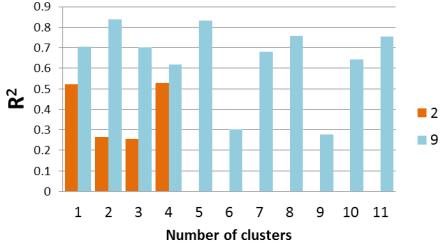, width=0.7\linewidth}
	\caption{\small Coefficient of determination ($R^2$) values for 2 and 9 friend clusters\label{fig:firstlast}}
\end{figure}

\begin{figure}
	\centering
		\epsfig{file=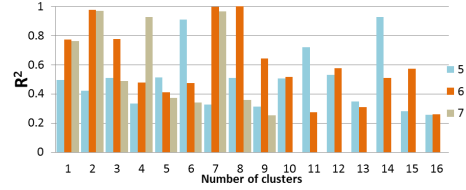, width=0.7\linewidth}
	\caption{\small Coefficient of determination ($R^2$) values for 5, 6 and 7 friend clusters \label{fig:clusterR2}}
\end{figure}

\textbf{Friend Clustering: }
In Figures \ref{fig:firstlast} and \ref{fig:clusterR2}, we show the adjusted coefficient of determination\footnote{The adjusted coefficient of determination is the proportion of variability in a data set that can be explained by the statistical model. This value shows how well future outcomes can be predicted by the model. $R^2$ can take 0 as minimum, and 1 as maximum.} ($R^2$) of our multiple regression model with different $k$ values for friend clustering. The x-axis gives the number of stranger clusters for which at least one friend cluster has an impact.  In Figure \ref{fig:firstlast} we see the performance for maximum and minimum number of friend clusters. For $k=2$, friend clusters are very roughly clustered, and each cluster is not homogeneous enough (i.e., contains different types of friends) to mine friend impacts\footnote{We use F-ratio probability to test the significance of parameters, i.e., a low probability (we use .05 as cutoff) for the F-ratio suggests that at least some of the friend cluster impacts are significant.}. As a result, we can observe friend impacts on very few clusters. For $k=9$, friend clusters are more homogeneous, but in this case our multiple regression model does not have many data points (strangers) to learn the impacts of friend clusters. 

Figure \ref{fig:clusterR2} shows the results for $k=5,6,7$ values. For two $k$ values, 5 and 6, we have the best results.  Our model hence suggests that friends of social network users can be put into 5 or 6 clusters when considering how much they can affect user decisions on stranger labeling.

\begin{figure}
	\centering
		\epsfig{file=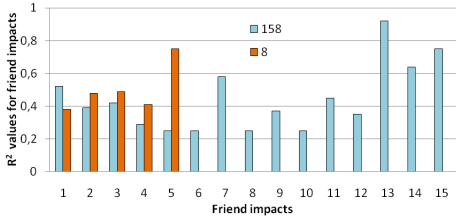, width=0.7\linewidth}
	\caption{\small Coefficient of determination ($R^2$) values of friend impacts for 158 and 8 stranger clusters. \label{fig:r2bycluster}}
\end{figure}

\begin{figure}
	\centering
		\epsfig{file=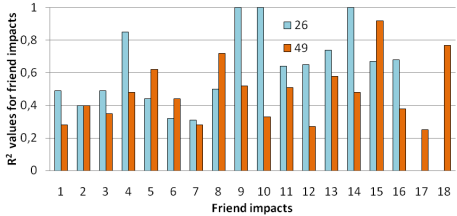, width=0.7\linewidth}
	\caption{\small Coefficient of determination ($R^2$) values of friend impacts for 26 and 49 stranger clusters \label{fig:r2byclustersfull}}
\end{figure}

\textbf{Stranger Clustering:}
In Figure \ref{fig:r2bycluster} we show how the $R^2$ values change for the biggest and smallest numbers of stranger clusters. With 8 stranger clusters, our model can detect friend cluster impacts on 5 out of 8 stranger clusters only, whereas for 158 clusters the number is 15 out of 158. For 158 stranger clusters, $R^2$ values are generally low because strangers are distributed into too many clusters, and each stranger cluster does not have many data points (strangers) to learn from. Although finding impacts on 5 out of 8 stranger clusters seems like a good performance, low $R^2$ values (lower than 0.5) show that the model can explain less than 50\% of the variation in data. In Figure \ref{fig:r2byclustersfull} we see that more stranger clusters can improve the model performance and this leads to $R^2$ values close to 1. For 26 stranger clusters, $R^2$ values are better, and we can find friend impacts in 16 out of 26 stranger clusters.

\textbf{Cross Validation: }
A major point in statistical modeling is the response to out of sample validation; a statistical model can be over-fitted to the training data, and it can perform poorly when applied to new testing data. After clustering and prior to learning friend cluster impacts, we prepare a test set for validating our model. We remove 10\% of strangers from stranger clusters and set those aside as the test strangers ($T$). Once friend impacts are found for stranger clusters, we plug in the set of test strangers, and calculate the root mean square value (RMSE) of their labels. RMSE for a stranger $s$ and user $u$ is defined by using the predicted label $\hat{L}_{us}$ and user given label $L_{us}$ as {$RMSE = \sqrt{\frac{\sum\limits_{s\in T}{(L_{us}-\hat{L}_{us})}}{\left|T\right|}}$}.

Cross validation results for different numbers of stranger clusters is detailed in Table \ref{table:clusterperformance} by using 6 friend clusters. The first row of the table shows the number of stranger clusters, whereas the second row shows the average $R^2$ values in these clusters. In the third row, we show the median size of stranger clusters; with increasing numbers of clusters, the number of strangers in each cluster decreases. In the case of 158, the average number of strangers in a cluster is reduced to 7, and this results in a poor performance because the model cannot have enough data to learn friend impacts on stranger clusters. The average number of validation points are shown in the fourth row. An increasing number of stranger clusters results in fewer validation points because some clusters will have less than 10 strangers themselves. In the fifth row, the root mean square values (RMSE) are shown for these validation points. In 26 stranger cluster our model yields the best $R^2$ and $RMSE$ pair results. 

These experimental results suggest that the optimal number of stranger clusters (26) is bigger than the optimal number of friend clusters ($k=5,6$). We explain this by the fact that although users can choose friends of specific characteristics, they cannot do so with strangers. As a result, strangers are more diverse than friends, and they need to be clustered differently from friends.

\begin{table}
\caption{ \small{Performance values for different numbers of stranger clusters.}}
	\label{table:clusterperformance}
\centering
  \begin{tabular}{|l | c  c c c c|}
  \hline
				Cluster count &8&26&49&82&158\\ 
			\hline
				$R^2$ &0.51&0.64&0.48&0.54&0.45\\
			\hline	
			Median Size &62&25&16&12&7\\
			\hline
			Validation points&179&99&69&48&27\\
			\hline
			RMSE & 0.35&0.45&0.62&0.97&0.94\\
			\hline
				\end{tabular}
	
\end{table}

\subsection{Friend Impacts and Risk Labels}
\label{sec:FriendImpactEstimates}

In this section we will give computed friend cluster impacts, and show how friends are assigned risk labels.

\begin{figure*}
\centering
\subfigure[5 friend clusters]{
\includegraphics[scale=0.3]{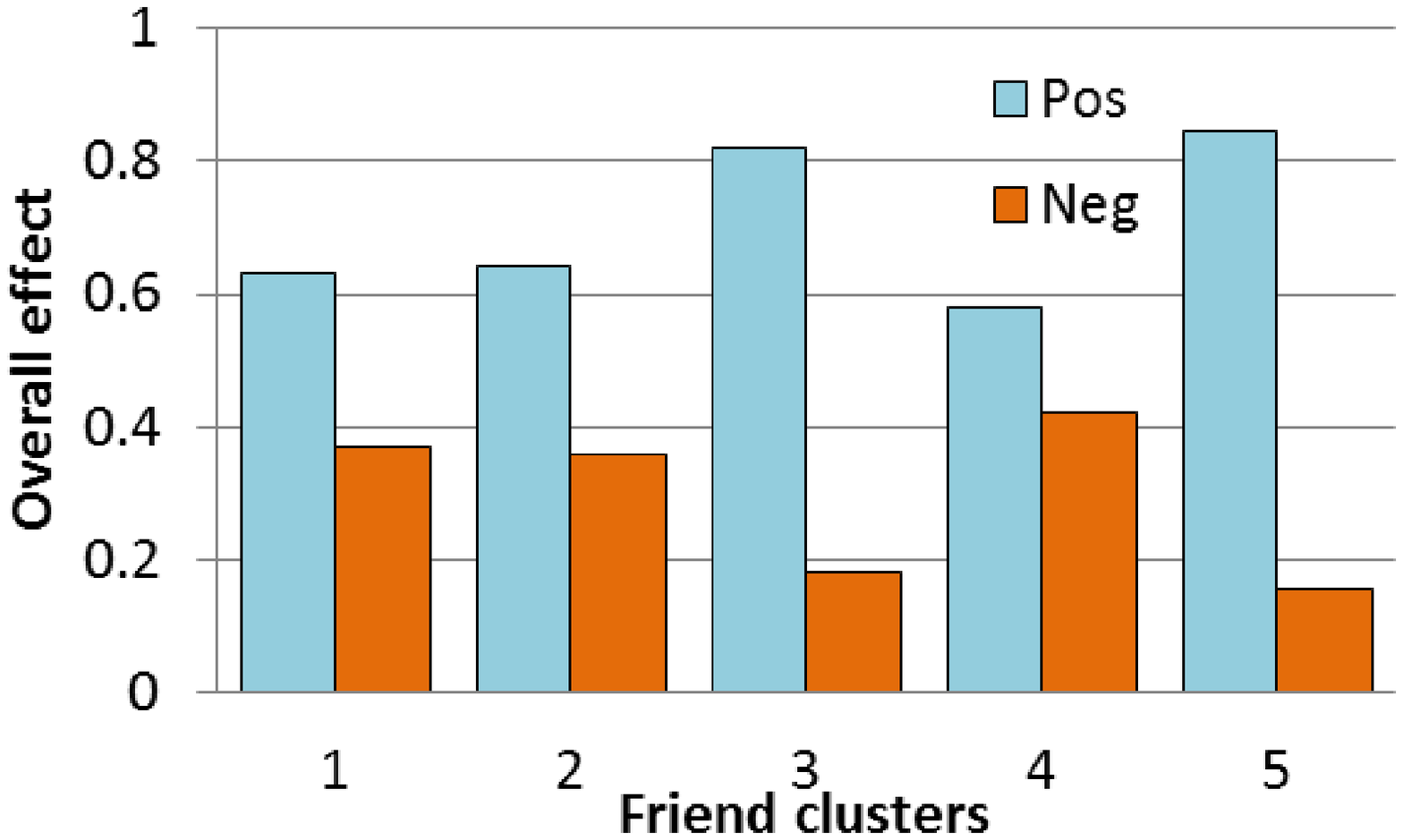}
\label{fig:frim5}
} 
\subfigure[6 friend clusters]{
\includegraphics[scale=0.3]{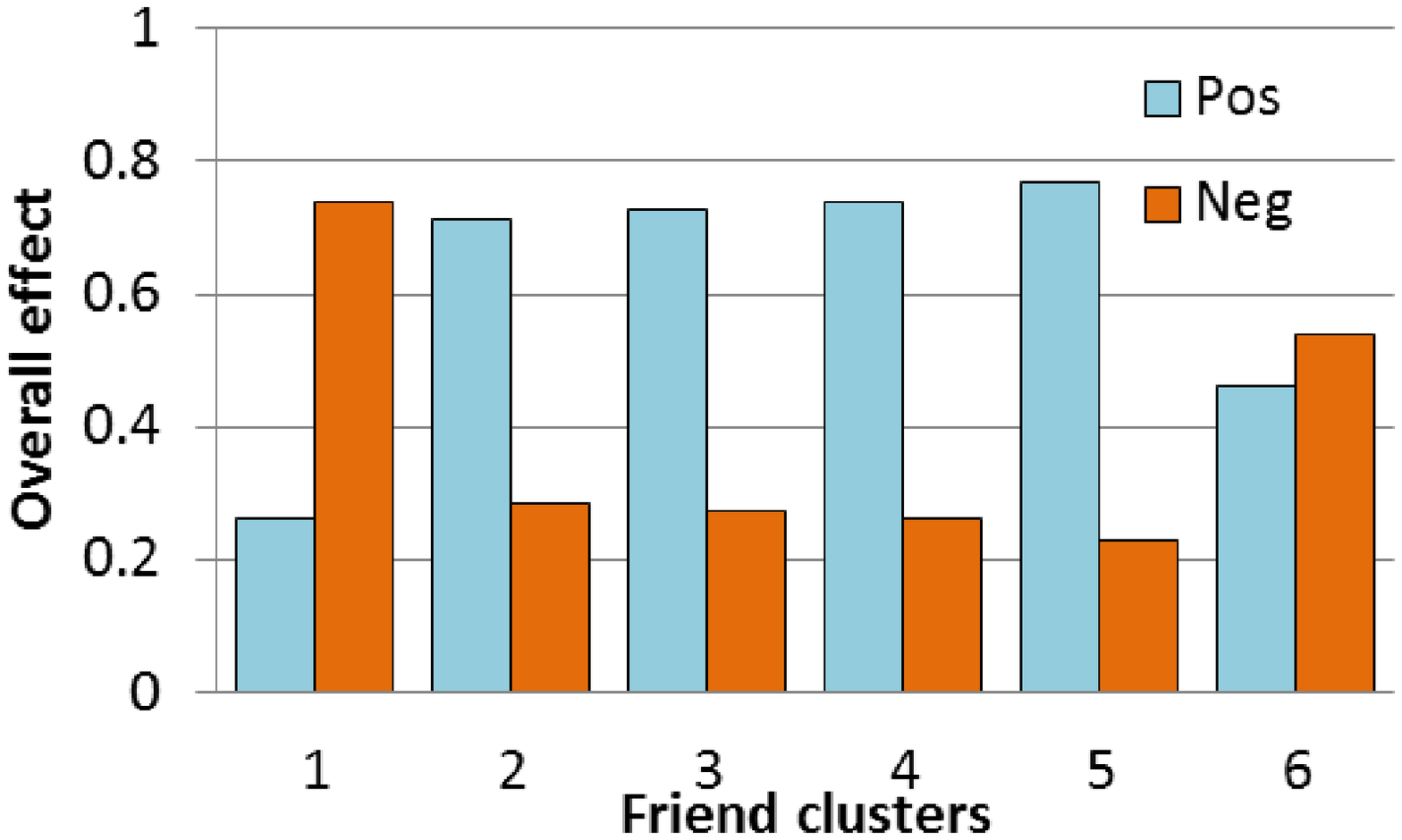}
\label{fig:frim6}
} 
\subfigure[7 friend clusters]{
\includegraphics[scale=0.3]{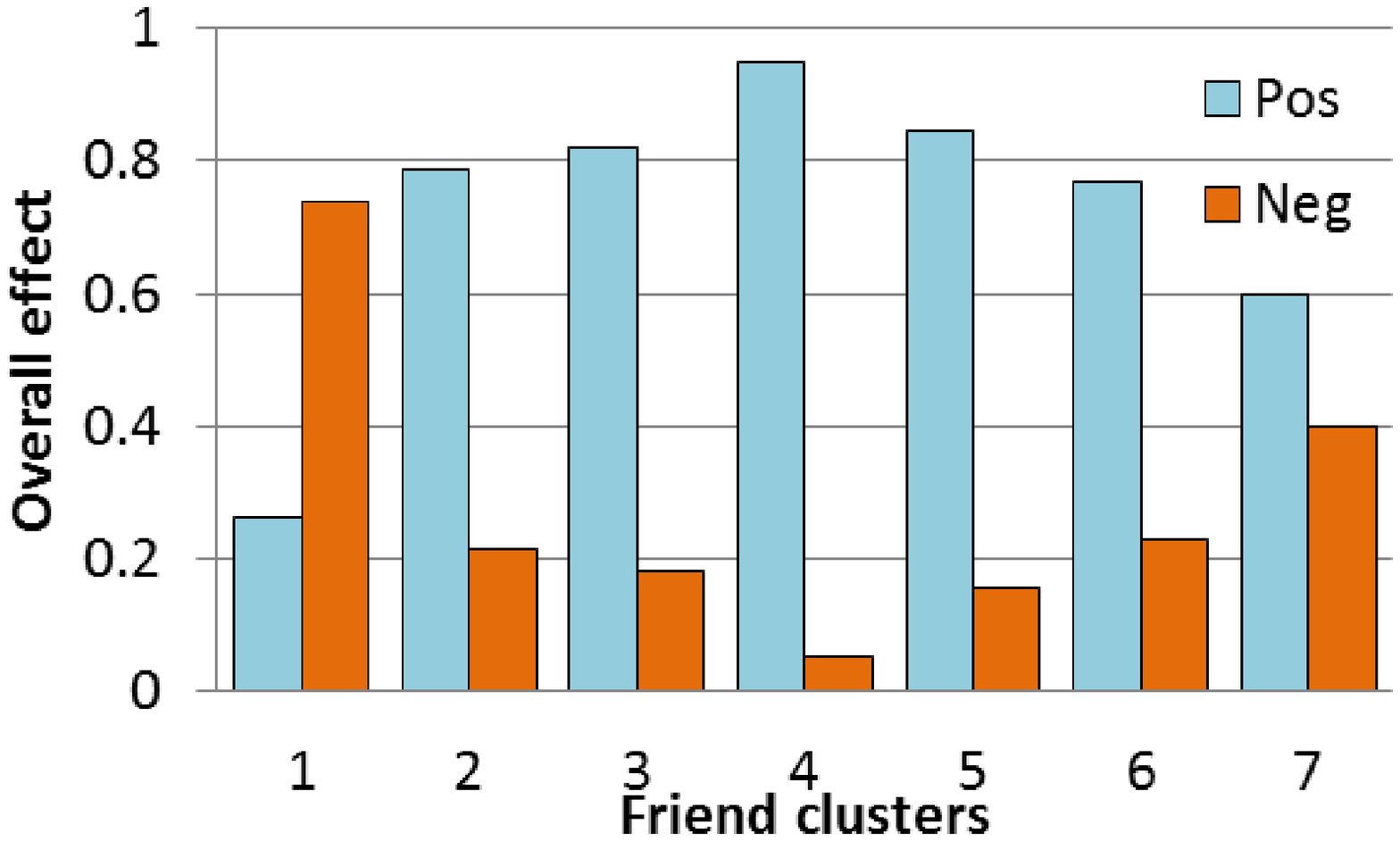}
\label{fig:frim7}
} 
\caption{Percentage of positive and negative impact values for friend clusters.}
\label{fig:frimp}
\end{figure*}

The rationale behind clustering was to observe different friend cluster impacts on different stranger clusters. Although a friend cluster can have an overall positive impact (i.e., reduces the risk label of most strangers), friend clusters might have different signs and multitudes of impact values on stranger clusters. In Figure \ref{fig:frimp} we show how different friend clusters can have positive and negative impact values for different $k$ values (number of friend clusters). Note that clusters are not identical across these figures, i.e., cluster 1 can have different members in each figure. This is because with different number of final clusters, the clustering algorithms produce potentially different clusters of data points. As seen in Figure \ref{fig:frim5}, when we increase the number of friend clusters from $k=5$ to $k=6$, positive and negative impact frequencies change for each cluster because either friend clusters became more homogeneous or some clusters did not have enough data points to learn from. Figure \ref{fig:frim6} shows two friend clusters with overall negative impacts (friend clusters 1 and 6). Figure \ref{fig:frim7} shows the positive and negative impact frequencies for $k=7$, where frequencies are more emphasized for negative and positive impacts of a cluster. Note that the number of overall negative clusters is reduced from 2 to 1 here. Similar to a transition from 5 to 6 clusters, friends of two negative clusters might be put into the same cluster (cluster 1) or  there were no longer enough strangers for some friend clusters to learn a negative impact. 

The existence of both positive and negative impact values for each friend cluster confirms our intuition that impacts of friend clusters vary depending on a stranger cluster. A friend is assigned a higher risk label when a friend cluster has a big percentage of negative impact values. In Section \ref{sec:FriendRisklabels}, we gave definitions of friend risk labels according to two threshold values (x=20, y=50) of negative impact percentages. By using k=6 friend clusters, from Figure \ref{fig:frim6} we see that friends from friend clusters 1 and 6 are labeled as very risky because the negative impact percentages for the clusters are $>0.6$. In the figure, we also see that none of the clusters have $<0.2$ negative impacts, hence no friends cluster is said to be not risky (label 1). 

We tested the accuracy of our risk definition for friends by observing 261 deleted friendships of users. As a performance measure, we assumed that the deleted friends should come from friends who are labeled as very risky, i.e., friends who belong to the 1st and 6th clusters. We have found 117 of the 261 deleted friends were found to belong to the 1st and 6th friend clusters. 

Although we chose to use specific values for very risky and not risky label thresholds (x=20, y=50) in assigning risk labels to strangers, our model can ask social network users to define these threshold values on their own. With this approach, our risk model for friends can be personalized by users and applied to privacy settings on social networks.

\section{Conclusion and Future Work}
\label{sec:Conclusion}

In this work, we looked into risks of friendships and analyzed how the risk labels of friends of friends can be used to compute risk labels of friends. We found that the number of mutual friends is not very important to change the risk perception of a user towards a friend of friend. On the other hand, having different types of mutual friends (i.e., friends from different friend clusters) with a friend of friend plays a bigger role in users' risk perception. Our results showed that in terms of risk, friends can be grouped into 6-7 clusters, whereas the number of groups for strangers can reach 26 or more. These results show that even though user numbers reach millions, friends for each user have similar roles. We have validated risk labels of friends on deleted Facebook friendships, and showed that risks of friendships can indeed be learned by considering users' risk perception towards friends of friends. In the future, we want to  create sets of global privacy settings by using our risk model, so that privacy settings can be automatically applied to different social network users.

\bibliographystyle{abbrv}
\bibliography{icdm} 
\end{document}